\documentclass[preprint,aps,showpacs,preprintnumbers,amsmath,amssymb]{revtex4}

\usepackage{epsfig}
\usepackage{bm}

\begin{document}

\title{The Deuteron as a Canonically Quantized Biskyrmion}

\author{A.~Acus$^{1,2}$ J.~Matuzas$^1$ E.~Norvai\v{s}as$^{1,2}$ and D.O.~Riska$^{3,4}$}
\affiliation{$^1$Institute of Theoretical Physics and Astronomy, Vilnius 2600 Lithuania\\
$^2$Department of Physics and Technology, Vilnius Pedagogical University, 2600 Lithuania\\
$^3$Department of Physical Sciences, University of Helsinki, 00014 Finland\\
$^4$Helsinki Institute of Physics, University of Helsinki, 00014 Finland}

\date{01.07.2003}
             
\begin{abstract}
The ground state configurations of the solution to Skyrme's
topological soliton model for systems with baryon number larger
than 1 are well approximated with rational map ans\"atze, without
individual baryon coordinates. Here the canonical quantization
of the baryon number 2 system,  
which represents the deuteron, is carried out in the rational
map approximation. The solution, which is described by the
six parameters of the chiral group SU(2)$\times$SU(2), is
stabilized by the quantum corrections. The matter density of the
variational quantized solution has
the required exponential large distance falloff and
the quantum numbers of the deuteron.
Similarly to the axially symmetric semiclassical solution, the
radius and quadrupole moment are, however, only about half 
as large as the corresponding
empirical values. The quantized deuteron solution is constructed for 
representations
of arbitrary dimension of the chiral group.
\end{abstract}


\maketitle

\section{Introduction}
The classical ground state solutions to Skyrme's topological
soliton model for baryons \cite{skyrme}, with baryon number 
($B$) larger than 1, have intriguing geometrical structure, 
with polyhedral symmetry \cite{Battye97}. The simplest example is the 
system with $B=2$, which has axial symmetry \cite{Manton97},
in agreement 
with the description of the deuteron based on a quantum
mechanical Hamiltonian for the interacting two-nucleon system 
\cite{Sick2001}.
Simple rational map ans\"atze, which provide remarkably 
accurate approximations to the classical ground state
configurations, have been derived for several systems with
baryon number larger than 1 \cite{Houghton98}. These rational
maps represent formal generalizations of Skyrme's
hedgehog ansatz for the single baryon. Here the 
rational map ansatz for the $B=2$ skyrmion
is employed to carry out the canonical quantization of
the $B=2$ solution, which represents the deuteron.

The Lagrangian density of the Skyrme model is chirally symmetric 
under constant SU(2)$\times$SU(2) transformations. 
The parameters of this symmetry group are treated as
the collective dynamical coordinates in the quantization
procedure.

The semiclassical quantization 
procedure developed in ref. \cite{Adkins83} for the nucleon 
employs only that half of the
parameters of full chiral symmetry group, which correspond to the 
diagonal subgroup, and therefore it describes only baryons 
with equal spin and isospin. 
The description by the SU(2) Skyrme model
of such states, which have unequal spin
and isospin as the deuteron, 
requires six dynamical 
variables in order to allow consideration of 
separate rotation of the generators of the SU(2) group and
the spatial vectors \cite{Braaten88}.

Because the generators are irreducible
tensors, these rotations are not independent, however, and 
therefore cannot be not be used for canonical
quantization. In the semiclassical quantization procedure the 
classical skyrmion is treated as a
rigid body with the consequence that the quantum mechanical rotation 
contributes only positive terms to the energy functional.
There is then no stable variational solution to the
quantized Hamiltonian. To obtain a stable variational
quantized solution one may draw on the canonical 
quantization procedure, which has been 
developed by K.~Fujii {\it et al.}~\cite{Fujii87} 
for the SU(2) Skyrme model. 
The treatment of the dynamical field 
variables of the Skyrme Lagrangian density as quantum-mechanical variables
{\it ab initio} generates negative quantum corrections 
and also stable quantum solitons~\cite{Acus98}. 
These energy of these quantum solitons depends on the dimension of 
the representation of the symmetry
group in contrast to the semiclassical case~\cite
{Norvaisas94,No97}.

The canonical quantization procedure developed below employs 
the two sets of three Euler angles, which correspond to left and right 
chiral rotation groups as the six collective coordinates. The 
resulting canonical angular momentum operators 
lead to compact forms for both the Lagrangian and Hamiltonian of the 
biskyrmion. The two sets of independent angular momentum operators allow 
construction of the eigenstate of the biskyrmion from the
eigenstates of two subsystems. For the deuteron the subsystems are the
neutron and proton, which form the state with common spin $S=1$ and
isospin $T=0$. The approach generalizes to dibaryons, which may be
constructed from neutrons and protons as well as from $\Delta$ 
resonances.

The matter density of the quantum soliton falls off exponentially
at long range in contrast to power law falloff of the classical
solution. The inverse of the length scale of this exponential falloff 
for the $B=1$ skyrmion corresponds to the pion mass~\cite{Acus98}.
In the case of the deuteron it should correspond to $2\sqrt{m_N B}$,
where $B$ is the binding energy and $m_N$ the nucleon mass.
It is shown here numerically in the rational map approximation 
that for the variational ground state the matter density
falls off at roughly this rate, as required. 

The approximate quantum soliton for the deuteron derived here describes
the rotational quantum corrections appropriately, but not the
large distance solution of two well separated single
skyrmions. This is revealed by the magnitude of its radius and
quadrupole moment, which are only about half as large as the 
corresponding empirical values. The semiclassical solution
shares these features \cite{Braaten88}.

The present manuscript is organized as follows. In Section~2 the classical
rational map ansatz for soliton with baryon number $2$ is generalized to 
representations of arbitrary dimension. In Section~3 this soliton (biskyrmion) is 
canonically
quantized with six collective variables, which correspond to the parameters of
chiral symmetry SU(2)$\times$SU(2) group. The expressions for the
electric form factor, quadrupole moment and rms radius
of the deuteron are presented in
Section~4. The numerical results for deuteron observables 
are discussed in Section~5.

\section{The classical axisymmetric soliton}

The Skyrme model is a Lagrangian density for a unitary
field $U(\bm{x},t)$, which may described by any representation of 
the $SU(2)$ group. In a general reducible representation 
the most compact expression of the unitary field 
$U({\bf x},t)$ is as a direct sum of Wigner's $D$ matrices for 
the irreducible representations of arbitrary integer or half
integer dimension $j$ \cite{Norvaisas94}:
\begin{equation}
U(\bm{x},t)=\sum_j \oplus D^j(\bm{\alpha} (\bm{x},t)).
\label{a1}
\end{equation}
The $D$ matrices depend on three unconstrained Euler angles 
$\bm{\alpha}=(\alpha ^1,\alpha ^2,\alpha ^3)$.

The chirally symmetric Lagrangian density has the form
\begin{equation}
{\cal L}[U(\bm{x},t)]=-{\frac{f_\pi ^2}4}\mathrm{Tr}\{R_\mu R^\mu \}+
{\frac 1{32e^2}}\mathrm{Tr}\{[R_\mu ,R_\nu ]^2\},
\label{a2}
\end{equation}
where the ''right'' current  is defined as
\begin{equation}
R_\mu =(\partial _\mu U)U^{\dagger },
\label{a3}
\end{equation}
and $f_\pi $ (the pion decay constant) and $e$ are parameters. 

The static variational solutions to classical Skyrme model
for baryon number $B=2$ (biskyrmion) have been derived numerically 
in refs.\cite{Manton97,Braaten88}. In 
ref.\cite{Houghton98} the  
following simple rational map ansatz, which preserves 
the axial symmetry of ground state solution for the biskyrmion,
was found to give an approximation to the ground state
energy, with an accuracy of better than 3 per cent:
\begin{equation}
\mathrm{e}^{\mathrm{i}\left(\hat{\bm{n}}\cdot \bm{\sigma}\right)
F_R(r)}\Longrightarrow 
U_{R}(\bm{r})={\mathrm{exp}}\{\mathrm{i}\hat{n}^{a}\cdot \hat{J}_{a}F_R(r)\}.  
\label{a4}
\end{equation}
Here $\hat{J}_{a}$ are SU(2) generators, which may be defined
in representations of arbitrary dimension. 
The scalar function $F_R(r)$ is the "chiral angle" for the
biskyrmion, which is determined by the variational equation
of motion. The circular components of the unit vector $\hat{\bm{n}}$
are
\begin{eqnarray}
\hat{n}_{+1}&&=-\hat{n}^{-1}=-\frac{\sin^2\vartheta}{\sqrt{2}
\left(1+\cos^2\vartheta\right)}\mathrm{e}^{2\mathrm{i}\varphi},\notag\\
\hat{n}_{0}&&=\hat{n}^{0}=\frac{2\cos\vartheta}{1+\cos^2\vartheta}
\label{a5},\\
\hat{n}_{-1}&&=-\hat{n}^{+1}=\frac{\sin^2\vartheta}{\sqrt{2}
\left(1+\cos^2\vartheta\right)}\mathrm{e}^{-2\mathrm{i}\varphi}.\notag
\end{eqnarray}

Substitution of the ansatz (\ref{a4}) in the Lagrangian density (\ref{a2}) 
yields the following expression for the mass density
of the classical biskyrmion:
\begin{eqnarray}
\mathcal{M}_{cl}=&&\frac{N}{2}\Bigl\{f_\pi^2\bigl[ F_R^{\prime 2}(r)
+2\mathcal{I} \sin^2F_R(r)\bigr]
\nonumber\\
&&+\frac{2\mathcal{I}}{e^2}\sin^2F_R(r)\bigl[ F_R^{\prime 2}
(r)+\frac{\mathcal{I}}{2}\sin^2F_R(r)\bigr] \Bigr\}.
\label{a6}
\end{eqnarray}
Here $\mathcal{I}$ is the Gaussian curvature, defined as
\begin{equation}
\mathcal{I} =\frac{4\sin^2\vartheta}{r^2(1+\cos^2\vartheta)^2}.
\end{equation}
In contrast to the hedgehog ansatz for $B=1$ this mass density depends on
both the polar angle $\vartheta$ and the radius $r$. 
The classical Lagrangian density depends on representation only 
through the overall factor $N=\frac23 \sum_j j(j+1)(2j+1)$,
where $2j+1$ is the dimension of the representation, and
which may be absorbed by renormalization of the model 
parameters \cite{Norvaisas94}. 

The requirement that the soliton mass be stationary yields the 
following differential 
equation for the chiral angle:
\begin{alignat}{2}
F_R^{\prime\prime}(\tilde r)\left(\tilde r^2+4 
\sin^2{F_R(\tilde r)}\right)&+
2F_R^{\prime 2}(\tilde r) \sin{2F_R(\tilde r)}+2 
\tilde r F_R^\prime(\tilde r)\notag\\
&-\sin2F_R(\tilde r) \bigl(2 {\tilde r^2}+
\bigl(\frac83+\pi\bigr)\sin^2F_R(\tilde r)
\bigr)=0.
\label{a7}
\end{alignat}
Here the dimensionless variable $\tilde r$ is defined as 
$\tilde r=ef_\pi r$. At large distances $\tilde r\rightarrow\infty$ 
this equation reduces to the asymptotic form
\begin{equation}
\tilde r^2 F_R^{\prime\prime}(\tilde r)+2 \tilde r F_R^\prime(\tilde r)
- 4F_R(\tilde r)=0.
\label{a8}
\end{equation}
The solution to the asymptotic equation (\ref{a8}) falls of with
an algebraic power of distance:
\begin{equation}
F_R(\tilde r)=C\tilde r^{-\frac{1+\sqrt{17}}{2}}.
\label{a9}
\end{equation}
The falloff rate is somewhat larger for the biskyrmion than 
for the hedgehog ansatz for $B=1$, 
as the power of $\tilde r$ in Eq.~(\ref{a9}) is $-2.56$ , whereas in
the case 
of $B=1$ it is $-2$. After renormalization by the 
factor $N$, the biskyrmion baryon 
density takes the form
\begin{equation}
\mathcal{B}^0 (r)=-\frac{\mathcal{I}}{2\pi ^{2}}F_{R}^{\prime }(r)\sin ^{2}F_{R}(r).
\label{a10}
\end{equation}

\section{Canonical quantization with six collective variables}

The Skyrme Lagrangian (\ref{a2}) is symmetric under 
chiral SU(2)$\times$SU(2) transformations. The canonical quantization
of the classical soliton solution (\ref{a7}) can be achieved by means of 
collective coordinates that separate the time dependent variables
from those that depend on the spatial coordinates:
\begin{equation}
U(\bm{x},\bm{\alpha}(t),\bm{\beta}(t))=A\left(\bm{\alpha}(t)\right) 
U_R(\hat{\bm{n}},F_R(r)) B^\dagger \left(\bm{\beta}(t)\right).
\label{bb1}
\end{equation}
Here the two sets of three Euler angles 
$\bm{\alpha}(t)=\bigl(\alpha ^1(t),\alpha 
^2(t),\alpha ^3(t)\bigr)$ and $ \bm{\beta}(t)=\bigl(\beta ^1(t),\beta ^2(t),
\beta ^3(t)\bigr)$ are those for the two SU(2) groups respectively.
In the canonical quantization
the Skyrme model is considered quantum mechanically {\it ab initio}. 
The collective coordinates $\bm{\alpha}(t),\bm{\beta}(t)$ and 
velocities $\dot{\bm{\alpha}}(t), \dot{\bm{\beta}}(t)$ are treated
as dynamical variables with the commutation relations
\begin{eqnarray}
\label{commutrel}
\left[ \dot{\alpha}^{k},\alpha ^{l}\right] &=&-\mathrm{i}\, {}_{R}f^{kl}(\bm{\alpha} ,\bm{\alpha}),
\qquad 
\left[ \dot{\alpha}^{k},\beta ^{l}\right] =-\mathrm{i}\, {}_{R}f^{kl}(\bm{\alpha},\bm{\beta} ), \\
\left[ \dot{\beta}^{k},\beta ^{l}\right] &=&-\mathrm{i}\, {}_{R}f^{kl}(\bm{\beta} ,\bm{\beta}),
\qquad 
\left[ \dot{\beta}^{k},\alpha ^{l}\right] =-\mathrm{i}\, {}_{R}f^{kl}(\bm{\beta},\bm{\alpha} ).\notag
\end{eqnarray}
The functions ${}_{R}f^{kl}$ are defined in (\ref{b10}) below.

Substitution of ansatz (\ref{bb1}) into the Lagrangian density (\ref{a2})
yields the quantum Lagrangian, which is quadratic in the
generalized velocities:
\begin{alignat}{2}
\label{b3}
L &=&\frac{1}{2}\dot{\alpha}^{k}g_{kl}(\bm{\alpha} ,\bm{\alpha} )\dot{\alpha}^{l}
+\frac{1}{2}\dot{\alpha}^{k}g_{kl}(\bm{\alpha} ,\bm{\beta} )\dot{\beta}^{l}
+\frac{1}{2}\dot{\beta}^{k}g_{kl}(\bm{\beta} ,\bm{\alpha} )\dot{\alpha}^{l} \\
&&+\frac{1}{2}\dot{
\beta}^{k}g_{kl}(\bm{\beta} ,\bm{\beta} )\dot{\beta}^{l}
+[(\dot{\bm{\alpha}},\dot{\bm{\beta}})^0 {\rm -order\ term }].
\end{alignat}
Here the coefficients $g_{kl}$ are defined as
\begin{eqnarray}
g_{kl}(\bm{\alpha} ,\bm{\alpha} ) &=&C_{k}^{\prime (a)}(\bm{\alpha} )
{}_{1}R_{ab}(F_{R})C_{l}^{\prime (b)}(\bm{\alpha} ),\notag \\
g_{kl}(\bm{\beta} ,\bm{\beta} ) &=&C_{k}^{\prime (a)}(\bm{\beta} )
{}_{1}R_{ab}(F_{R})C_{l}^{\prime (b)}(\bm{\beta} ), \\
g_{kl}(\bm{\alpha} ,\bm{\beta} ) =g_{lk}(\bm{\beta} ,\bm{\alpha} )&=&C_{k}^{\prime (a)}(\bm{\alpha} )
{}_{2}R_{ab}(F_{R})C_{l}^{\prime (b)}(\bm{\beta} ).\notag
\label{b4}
\end{eqnarray}
The  $C_{k}^{\prime (a)}$'s and their inverses $C_{(a)}^{\prime k}$ 
are functions of the dynamical variables, which appear 
in the differentiation of the Wigner
$D$ matrices~\cite{Norvaisas94}:
\begin{equation}
\frac{\partial}{\partial \alpha^i} D_{mn}^j(\bm{\alpha} )=
-\frac 1{\sqrt{2}}\ C_i^{\prime (a)}(\bm{\alpha}
)\ D_{mm^{\prime }}^j(\bm{\alpha} )\Bigr\langle jm^{\prime }
\bigl| J_a\bigr| jn\Bigr\rangle .
\label{ccoef}
\end{equation}
The matrices ${}_{1,2}R_{ab}(F_{R})$ are antidiagonal

\begin{equation}
{}_{1}R_{ab}(F_{R})=(-1)^{a}{}\ _{1}^{a}R(F_{R})\delta _{a,-b},\qquad \
{}_{2}R_{ab}(F_{R})=(-1)^{a}{}\ _{2}^{a}R(F_{R})\delta _{a,-b},
\label{b5}
\end{equation}
and have the matrix elements
\begin{eqnarray}
\label{Rdef}
_{1}^{\pm }R(F_{R}) &=&-Y+\frac{\pi }{2e^{3}f_\pi}\int 
\mathrm{d}\tilde r \tilde r^2
\Bigl( \frac{\pi}{4} F_{R}^{\prime 2}+\frac{8}{3 \tilde r^2}
 \sin ^{2}F_{R}\Bigl),
\notag \\
_{1}^{0}R(F_{R}) &=&-Y-\frac{\pi }{2e^{3}f_\pi}\int \mathrm{d}\tilde r 
\tilde r^2 \Bigl( \frac12 (4-\pi)F_{R}^{\prime 2}+\frac{8}{3 \tilde r^2}
 \sin ^{2}F_{R}\Bigl),
\\
_{2}^{\pm }R(F_{R}) &=&Y-\frac{\pi }{4e^{3}f_\pi}\int \mathrm{d}\tilde r 
\tilde r^2 \Bigl( \pi\sin ^{2}F_{R} +\frac12  \pi \cos 2 F_{R} 
F_{R}^{\prime 2}+\frac{4}{3\tilde r^2}\sin ^{2} 2 F_{R} \Bigl),
\notag \\
_{2}^{0}R(F_{R}) &=&Y-\frac{\pi }{4e^{3}f_\pi}\int \mathrm{d}\tilde r 
\tilde r^2 \Bigl( (4-\pi)\sin ^{2}F_{R} -(4-\pi )\cos 2 F_{R} F_{R}^{\prime 
2}\notag \\
&-&\frac{4}{3\tilde r^2}\sin ^{2} 2 F_{R} \Bigl).\notag
\end{eqnarray}
These matrix elements contain the infinite integral
\begin{equation}
Y=\frac{\pi }{2e^{3}f_\pi}\int \mathrm{d}\tilde r \tilde r^2,
\label{b7}
\end{equation}
which drops out from the generalized moments of inertia and the
mass density, when taken to infinite, but 
which is convenient to retain formally in the 
intermediate steps.

The infinite terms arises in the quadratic term in the 
Lagrangian density, when the left and right rotations are
unequal:
\begin{eqnarray}
\mathrm{Tr}\,R^0 R_0\approx &&\mathrm{Tr}\dot A A^{\dag }
\dot A A^{\dag }+ \mathrm{Tr}\,\dot B B^{\dag }\dot B B^{\dag } \notag \\
&&-\mathrm{Tr}\,\dot A A^{\dag }U_0 \dot B B^{\dag } U_0^\dag -
\mathrm{Tr}\,U_0 \dot A A^{\dag }U_0^\dag \dot B B^{\dag}. 
\label{infities}
\end{eqnarray}
Here only the terms, which contain $\dot{\bm{\alpha}}$ or 
$\dot{\bm{\beta}}$, and which are important for commutation relations
are considered. 
The infinite terms in (\ref{Rdef}) arise from 
the terms on the r.h.s. of (\ref{infities}),
which are independent of the
spatial coordinates.
In the case when $A=B$, $U_0\rightarrow 1$ when
$r\rightarrow\infty$ the infinities disappear
from the Lagrangian (\ref{b3}).

The Lagrangian (\ref{b3}) may be used to define the following
canonical momentum operators, which are conjugate to 
the collective coordinates:
\begin{eqnarray}
\pi _{k}(\bm{\alpha} ) &=&\frac{\partial L}{\partial \dot{\alpha}^{k}}=\frac{1}{2}
\left\{ \dot{\alpha}^{l},g_{kl}(\bm{\alpha} ,\bm{\alpha} )\right\} +\frac{1}{2}\left\{ 
\dot{\beta}^{l},g_{kl}(\bm{\alpha} ,\bm{\beta} )\right\}, \\
\pi _{k}(\bm{\beta} ) &=&\frac{\partial L}{\partial \dot{\beta}^{k}}=\frac{1}{2}
\left\{ \dot{\beta}^{l},g_{kl}(\bm{\beta} ,\bm{\beta} )\right\} +\frac{1}{2}\left\{ 
\dot{\alpha}^{l},g_{kl}(\bm{\beta} ,\bm{\alpha} )\right\}.\notag
\label{b8}
\end{eqnarray}
Here the curly brackets denote anticommutators. The canonical 
commutation relations
\begin{equation}
\bigl[ \pi _{k}(\bm{\alpha} ),\alpha ^{l}\bigr]=-i\delta _{k,l},\qquad \bigl[ \pi
_{k}(\bm{\beta} ),\beta ^{l}\bigr]=-i\delta _{k,l},\qquad \bigl[ \pi _{k}(\bm{\alpha}
),\beta ^{l}\bigr]=\bigl[\pi _{k}(\bm{\beta} ),\alpha ^{l}\bigr]=0
\label{b9}
\end{equation}
lead to the system of linear equations for the functions 
$_Rf^{kl}$ in (\ref{commutrel}),  
the solution of which can be written in the form
\begin{eqnarray}
\label{b10}
{}_{R}f^{kl}(\bm{\alpha} ,\bm{\alpha} ) &=&C_{(a)}^{\prime k}(\bm{\alpha})
{}_{1}F^{ab}(F_{R})C_{(b)}^{\prime l}(\bm{\alpha} ), \notag \\
{}_{R}f^{kl}(\bm{\beta} ,\bm{\beta} ) &=&C_{(a)}^{\prime k}(\bm{\beta})
{}_{1}F^{ab}(F_{R})C_{(b)}^{\prime l}(\bm{\beta} ), \\
{}_{R}f^{kl}(\bm{\alpha} ,\bm{\beta} ) &=&_{R}f^{lk}(\bm{\beta} ,\bm{\alpha} )=C_{(a)}^{\prime k}
(\bm{\alpha} ){}_{2}F^{ab}(F_{R})C_{(b)}^{\prime l}(\bm{\beta} ).\notag
\end{eqnarray}
Here the antidiagonal matrices
\begin{eqnarray}
\label{b11}
{}_{1}F^{ab}(F_{R}) &=&(-1)^{a}{}\ _{1}^{a}F(F_{R})\delta _{a,-b}, \\
_{2}F^{ab}(F_{R}) &=&(-1)^{a}\ {}_{2}^{a}F(F_{R})\delta _{a,-b},\notag
\end{eqnarray}
have the following matrix elements, which in the 
limit $Y\longrightarrow\infty$ become finite:
\begin{eqnarray}
\lim\limits_{Y\rightarrow \infty }{}_{1}^{\pm }F 
&=&\lim_{Y\rightarrow\infty}\frac{{}_{2}^{\pm }R}{({}_{2}^{\pm }R)^{2}
-({}_{1}^{\pm }R)^{2}}
=-\frac{2}{a_{1}},\notag \\
 \lim\limits_{x\rightarrow \infty }{}_{2}^{\pm }F
&=&\lim_{Y\rightarrow\infty}\frac{-{}_{1}^{\pm }R}{({}_{2}^{\pm}R)^{2}
-({}_{1}^{\pm }R)^{2}}=-\frac{2}{a_{1}}, \notag \\
\lim\limits_{Y\rightarrow \infty }{}_{1}^{0}F 
&=&\lim\limits_{Y\rightarrow \infty }\frac{_{2}^{0}R}
{({}_{2}^{0}R)^{2}-(_{1}^{0}R)^{2}}=-\frac{2}{a_{0}},
\notag \\
\lim\limits_{Y\rightarrow \infty }{}_{2}^{0}F &=&
\lim\limits_{Y\rightarrow \infty }\frac{- {}_{1}^{0}R}
{({}_{2}^{0}R)^{2}-({}_{1}^{0}R)^{2}}=
-\frac{2}{a_{0}}.
\label{b13}
\end{eqnarray}
The quantities
\begin{subequations}
\begin{align}
&a_0=\frac{{\tilde a}_0}{e^3f_\pi}=
\frac{2\pi}{e^3f_\pi}\int^{\infty}_{0}\mathrm{d}\tilde r\tilde r^2
\sin^2F_R\bigl((4-\pi)(1+F_R^{\prime 2})+8\frac{\sin^2 F_R}{3\tilde r^2}\bigr),
\label{b55}\\
&a_1=\frac{{\tilde a}_1}{e^3f_\pi}=
\frac{\pi}{e^3f_\pi}\int^{\infty}_{0}\mathrm{d}\tilde r\tilde r^2
\sin^2F_R\bigl(\pi(1+F_R^{\prime 2})+16\frac{\sin^2F_R}{3\tilde r^2}\bigr),
\label{b6}
\end{align}
\end{subequations}
define two different soliton moments of inertia, as appropriate for an 
axially 
symmetric system. It is convenient to 
introduce the following angular momentum operators
on on the hypersphere $S^3$, which is the group manifold of SU(2):
\begin{eqnarray}
\widehat{J}_{a}^{\prime }(\bm{\alpha} ) &=&-\frac{\mathrm{i}}{\sqrt{2}}\bigl\{ \pi_{k}
(\bm{\alpha} ),C_{(a)}^{\prime k}(\bm{\alpha} )\bigr\} , \\
\widehat{J}_{a}^{\prime }(\bm{\beta} ) &=&-\frac{\mathrm{i}}{\sqrt{2}}\bigl\{ \pi_{k}
(\bm{\beta} ),C_{(a)}^{\prime k}(\bm{\beta} )\bigr\},\notag
\label{bb15}
\end{eqnarray}
the components of which satisfy the standard commutation relations and
\begin{equation}
\bigl[ \widehat{J}_{a}^{\prime }(\bm{\alpha} ),\widehat{J}_{b}^{\prime }(\bm{\beta} )
\bigr] =0.
\label{b16}
\end{equation}
The coefficients of the quantized Lagrangian (\ref{b3}) contains the 
infinite integrals $Y$. After replacement of the velocities by the natural 
angular momentum operators (\ref{bb15}), the Lagrangian density can be 
reexpressed as a sum of the  angular momentum operators 
$\hat J^\prime (\bm{\alpha} ) +\hat J^\prime (\bm{\beta} )$. By means of some 
lengthy manipulation the Lagrangian density takes the following
form in terms of these:
\begin{eqnarray}
&&\mathcal{L}(\bm{\alpha} ,\bm{\beta} , \hat{\boldsymbol{n}} ,F_R(\tilde r)) =-\mathcal{M}_{cl}+
\frac12 f_{\pi }^{4} e^6 \sin ^{2}F_{R} \bigl( 1+ F^{\prime 2}_{R}+
\mathcal{I}\sin ^{2}F_{R}\bigr) \notag \\
&&
\times \biggl\{\frac{1}{\tilde a_{1}^{2}}
\bigl( \hat{J}^{\prime}(\bm{\alpha} )+\hat{J}^{\prime }(\bm{\beta} )\bigr) ^2+
\bigl( \frac{1}{\tilde a_{0}^{2}}-\frac{1}{\tilde a_{1}^{2}} \bigr) 
\bigl( \hat{J}_0^{\prime}(\bm{\alpha} )+\hat{J}_0^{\prime }(\bm{\beta} )\bigr) ^2 \\
&&-\Bigl[ \frac{1}{\tilde a_{1}}
\bigl(\hat{J}^{\prime}(\bm{\alpha} )+\hat{J}^{\prime }(\bm{\beta} )\bigr)\cdot 
\hat{\boldsymbol{n}}\bigr)
+\bigl(\frac{1}{\tilde a_{0}}-\frac{1}{\tilde a_{1}} \bigl)
\bigl( \hat{J}_0^{\prime}(\bm{\alpha} )+\hat{J}_0^{\prime }(\bm{\beta} )\bigr)\hat n_0
\Bigr] ^2 \biggr\}+ \Delta \mathcal{M}. \notag
\label{fullqL}
\end{eqnarray}
The last term on the r.h.s., $\Delta \mathcal{M}$, is the quantum correction to classical 
mass density, which appears on account of the commutation 
relation (\ref{b9}).
This has the expression
\begin{equation}
\Delta \mathcal{M}=f_{\pi }^{4} e^6\Bigl\{
\frac{\tilde {\mathcal{Q}}_3}{\tilde a^2_0}+\frac{\tilde {\mathcal{Q}}_4}{\tilde a_0 \tilde a_1}+
\frac{\tilde {\mathcal{Q}}_5}{\tilde a^2_1}+d \bigl( \frac{\tilde {\mathcal{Q}}_6}{\tilde a^2_0}+
\frac{\tilde {\mathcal{Q}}_7}{\tilde a_0 \tilde a_1}+
\frac{\tilde {\mathcal{Q}}_8}{\tilde a^2_1}\bigr)\Bigr\},
\label{deltaM}
\end{equation}
where 
\begin{eqnarray}
\label{qdens}
\tilde {\mathcal{Q}}_3 &=& \frac18 \sin ^{2}F_{R} (1-\hat n^2_0)
(1+F^{\prime 2}_{R}+\mathcal{I} \sin ^{2}F_{R}),\notag \\
\tilde {\mathcal{Q}}_4&=& \frac14 \sin ^{2}F_{R} (1-\hat n^2_0)
(1+F^{\prime 2}_{R}),\notag \\
\tilde {\mathcal{Q}}_5 &=& \frac18 \sin ^{2}F_{R}
 \bigl[ (1+3\hat n^2_0)(1+F^{\prime 2}_{R})+
(1+\hat n^2_0)\mathcal{I} \sin ^{2}F_{R}\bigr],\notag\\
\tilde {\mathcal{Q}}_6 &=& \frac{1}{5\cdot 8} \sin ^{2}F_{R} (1-\hat n^2_0)
\bigl[ (1-\hat n^2_0)(3\sin ^{2}F_{R}+ 3 F^{\prime 2}_{R}-2 F^{\prime 2}_{R}
\sin ^{2}F_{R})\\
&& + (1+3\hat n^2_0)\mathcal{I} \sin ^{2}F_{R}\bigr],\notag\\
\tilde {\mathcal{Q}}_7 &=& \frac{1}{4\cdot 5} \sin ^{2}F_{R} (1-\hat n^2_0)
\bigl[(1+3\hat n^2_0)(\sin ^{2}F_{R}+ F^{\prime 2}_{R})  
\notag \\
&&-\sin ^{2}F_{R}(2 (1+\hat n^2_0) F^{\prime 2}_{R} 
+3\mathcal{I} \hat n^2_0)\bigr],\notag\\
\tilde {\mathcal{Q}}_8 &=& \frac{1}{5\cdot 8} \sin ^{2}F_{R} 
\bigl[(3+2\hat n^2_0+3\hat n^4_0)(\sin ^{2}F_{R}+ F^{\prime 2}_{R}) - 
\sin ^{2}F_{R}(2 (1+\hat n^2_0) F^{\prime 2}_{R} \notag \\
&&-(1+4\hat n^2_0-3\hat n^4_0)\mathcal{I} )\bigr].\notag
\end{eqnarray}
Here only the quantum correction depends on the dimension of
the representation of the chiral 
field $U$ through the explicit factor
\begin{equation}
d=\frac32 \frac{\sum_j j(j+1)(2j+1)(2j-1)(2j+3)}{\sum_j j(j+1)(2j+1)}.
\end{equation}
Traditionally the Skyrme model is formulated in the fundamental 
representation, in which $j=\frac12$ and $d=0$. 

The operators in the Hamiltonian
of the biskyrmion system also are the sum of two independent angular momentum
operators $\hat{J}^\prime (\bm{\alpha})+\hat{J}^\prime (\bm{\beta})$. The terms
with the operators $\bigl(\hat{J}^\prime (\bm{\alpha})\bigr)^2$, 
$\bigl(\hat{J}^\prime (\bm{\beta})\bigr)^2$ or 
$\hat{J}^\prime (\bm{\alpha} ) \hat{J}^\prime (\bm{\beta} )$  
drop out from the Hamiltonian as they contain coefficients
with the infinite factor $Y$ in their denominators.
The angular momentum
operators are natural operators for Skyrme model and in terms of
them the Hamiltonian operator for the biskyrmion becomes:
\begin{eqnarray}
H(\bm{\alpha} ,\bm{\beta} ,F_R)=&&M_{cl} +\frac{1}{2 a_1} \bigl( \hat{J}^{\prime
}(\bm{\alpha} )+\hat{J}^{\prime }(\bm{\beta} )\bigr) ^2
\notag \\
&& +\bigl(\frac{1}{a_0^2
}-\frac{1}{a_1^2} \bigr) 
\bigl( \hat{J}_0^{\prime}(\bm{\alpha} )+\hat{J}_0^{\prime }(\bm{\beta} )\bigr) ^2 +
\Delta M,
\label{hamiltonian}
\end{eqnarray}
where 
\begin{equation}
\Delta M = \int \mathrm{d}^3 \bm{r} \Delta \mathcal{M},
\end{equation}
is the quantum correction to soliton mass. The Hamiltonian is similar
to semiclassically quantized Hamiltonian of a rotator, with exception for
the  quantum correction. The normalized eigenstate vectors for the
Hamilton 
operator (\ref{hamiltonian}) can be constructed from eigenstates of
two subsystems with common spin $S$ and isospin $T$ as:
\begin{eqnarray}
\left| 
\renewcommand{\arraystretch}{0.8}
\begin{array}{ll}
S&T \\ 
l_{1}& l_{2} \\ 
m_{s}& m_{t}
\end{array}
\right\rangle &=& \frac{\sqrt{(2l_{1}+1)(2l_{2}+1)}}{16\pi ^{2}}\notag
\\
&&\times
\sum_{m_{1},m_{2},m_{1}^{\prime},m_{2}^{\prime }}
\biggl[ 
\setlength{\arraycolsep}{2pt}
\begin{array}{ccc}
l_{1} & l_{2} & S \\ 
m_{1} & m_{2} & m_{s}
\end{array}
\biggr] \biggl[ 
\setlength{\arraycolsep}{2pt}
\begin{array}{ccc}
l_{1} & l_{2} & T \\ 
m_{1}^{\prime } & m_{2}^{\prime } & m_{t}
\end{array}
\biggr]\notag \\ 
&&D_{m_{1}m_{1}^{\prime }}^{\ell_1}(\bm{\alpha} )D_{m_{2}m_{2}^{\prime
}}^{\ell_2}(\bm{\beta} )\bigl| 0\bigr\rangle.
\notag \\
&&
\end{eqnarray}

The operators $\hat{J}^\prime (\bm{\alpha})$ and $\hat{J}^\prime (\bm{\beta})$ are
``right rotation'' operators for the Wigner matrices $D^{\ell_1}(\bm{\alpha} )$ 
and $D^{\ell_2}(\bm{\beta} )$. The biskyrmion with different $S$
and $T$ can now be constructed from 
states with the quantum numbers of the nucleons and the $\Delta$ resonances. The 
eigenvalue of Hamiltonian operator gives the mass of quantum biskyrmion 
as
\begin{equation}
M_d = M_{cl} +\Delta M +\frac12 \Big[\frac{1}{a_1} T (T+1)+
\Bigl( \frac{1}{a_0}- \frac{1}{a_1}\Bigr) m^2_t \Bigr],
\label{mass}
\end{equation}
which depends only on isospin $T$ and isospin projection $m_t$. For the
deuteron $T=0$, and the variation of the mass (\ref{mass}) gives the 
integro-differential equation for the chiral angle $F_R(r)$
\begin{equation}
\frac{\delta M_d}{\delta F_R} =0.
\label{fullE}
\end{equation}
This explicit expression for the deuteron is identical to the
corresponding equation for
 dibaryons~\cite{Krupovnickas2001}. At large distances the equation 
(\ref{fullE}) reduces to the asymptotic form
\begin{align}
&\tilde r^2F_R^{\prime\prime}+2\tilde 
rF_R^\prime-\left(4+\tilde m^2\tilde r^2\right)F_R=0,
\label{b150}
\end{align}
where
\begin{eqnarray}
\label{m2}
\tilde{m}^2&=&e^4 \Bigl\{ -\frac{4-\pi}{2 \tilde{a}_0^2} \bigl( m_t^2 +\frac14\bigr)
-\frac{1}{4 \tilde{a}_1^2}\Bigl[\bigl(T (T+1)-m_t^2+\frac32\bigr)\pi+2\Bigr]
-\frac{4-\pi}{4 \tilde{a}_0 \tilde{a}_1}\notag \\
&&+\frac{2(4-\pi)}{\tilde{a}_0^3}\tilde{Q}_3
+\Bigl(\frac{4-\pi}{\tilde{a}_0^2 \tilde{a}_1}
+\frac{\pi}{2\tilde{a}_0 \tilde{a}_1^2}\Bigr)\tilde{Q}_4
+\frac{\pi}{\tilde{a}_1^3}\tilde{Q}_5\\
&&+d \Bigl[\frac{2(4-\pi)}{\tilde{a}_0^3}\tilde{Q}_6
+\Bigl(\frac{4-\pi}{\tilde{a}_0^2\tilde{a}_1}
+\frac{\pi}{2\tilde{a}_0\tilde{a}_1^2}\Bigr)\tilde{Q}_7
+\frac{\pi}{\tilde{a}_1^3}\tilde{Q}_8\Bigr]\Bigr\},\notag
\label{mtilde}
\end{eqnarray}
and 
\begin{equation}
\tilde{Q}_k=\int\mathrm{d}^3 \bm{r} {\mathcal{Q}}_k .
\end{equation}
The factor $\tilde{m}$ describes the falloff rate of the chiral angle at 
large distances:
\begin{align}
&F_R\left(\tilde 
r\right)=C\Bigl(\frac{\tilde m}{\tilde r}+\frac{2}{\tilde r^2}\Bigr)
\mathrm{e}^{-\tilde m\tilde r}.
\label{b170}
\end{align}
The related quantity $m=ef_{\pi}\tilde m$ describes the asymptotic falloff
$\exp (-2mr)$ of biskyrmion mass density like Yukawa pion cloud for nucleon.

\section{Structure of the quantized deuteron solution}

The electric and quadrupole form factors $G_C(Q^2)$ and $G_Q(Q^2)$ 
of the deuteron state are obtained as the matrix elements of the 
spin-scalar and spin-tensor parts of the time component of the 
electromagnetic current operator. For the isospin $0$ deuteron
state this is given by the anomalous baryon current.
The matrix element is evaluated between the deuteron states in the 
Breit frame, which is defined by 
$\boldsymbol{p}+\boldsymbol{p}^\prime=0$ (\cite{Braaten88}):
\begin{eqnarray}
&&\bigl\langle d\ , m^\prime _s \, 
\boldsymbol{p}^\prime \bigr| J^0 (\boldsymbol{r}=0)
\bigl| d\, m _s \, \boldsymbol{p} \bigl\rangle = 
G_C(Q^2) \delta_{m_s\,m^\prime _s}
\notag \\
&&+\frac{1}{6 M^2_d} G_Q(Q^2) U_{m^\prime _s\, a} 
\bigl(3 q^a q^b-q^2 \delta_{a\,b}\bigr)
U^\dagger _{b\,m_s}.
\label{formfdef}
\end{eqnarray}
Here $\mathbf{q}=\mathbf{p}^\prime -\mathbf{p}$ is the momentum 
transfer, $Q^2=-q^2$, $M_d$ is the mass of the deuteron,
and $U_{m_s\, a}$ is the unitary matrix that relates the
Cartesian and spherical bases.

The expression for the electric form factor is:
\begin{equation}
G_C(Q^2)=\frac12 \int \mathrm{d}^3\mathbf{r} j_0(q r) \mathcal{B} ^0(\mathbf{r}),
\label{Coulombff}
\end{equation}
where $j_k(q r)$  is the spherical Bessel function of $k$-th order. 
The quadrupole form factor is correspondingly
\begin{equation}
G_Q(Q^2)=\frac32 \frac{M^2_d}{q^2}\int \mathrm{d}^3\mathbf{r}
 (1-3\cos^2 \theta ) j_2(q r) \mathcal{B} ^0(\mathbf{r}).
\label{quadrupoleff}
\end{equation}

The mean square charge radius and the quadrupole moment are
defined as
\begin{equation}
r^2= -6 \frac{\mathrm{d}}{\mathrm{d} Q^2} G_C(Q^2),\quad Q=M_d^2 G_Q(0).
\end{equation}  
It follows that
\begin{equation}
\langle r^2\rangle _{ch}=\frac12 \int \mathrm{d}^3\mathbf{r}
 r^2 \mathcal{B} ^0(\mathbf{r})
\label{radiusch}
\end{equation}
and that 
\begin{equation}
Q_d=\frac{1}{10} \int \mathrm{d}^3\mathbf{r} r^2 (1-3\cos^2 \theta )
 \mathcal{B} ^0(\mathbf{r}).
\label{quadrupolemomentum}
\end{equation}
The matter radius of the deuteron solution, $\langle r^2\rangle _m$, may
in turn be determined from deuteron mass distribution as
\begin{equation}
\langle r^2\rangle _m=\frac{1}{M_d} \int \mathrm{d}^3\mathbf{r} r
^2 \bigl(\mathcal{M} (\mathbf{r})+\Delta \mathcal{M}(r)\bigr) .
\label{massdistribution}
\end{equation}

\section{Numerical results and discussion}

The numerical value for the rate $\tilde m$ (\ref{m2}), at which the 
mass density of the solution decays with distance provides the key test 
of the phenomenological gain in the canonically quantization of
the ground state solution with the quantum numbers of the deuteron.
In order to
match the falloff rate of the matter density that corresponds to
the deuteron wave function, it should equal $2\sqrt{B M_n}$,
where $B$ is the binding energy and $M_n$ the nucleon mass.
The empirical value of this quantity is 91.4 MeV. This test 
requires numerical solution of the variational problem for the
quantized deuteron state. For this purpose the
two parameters $f_\pi$ and $e$ of the Lagrangian density
of the Skyrme model have, however, to be determined first
by fits to two empirical nucleon observables. 

The procedure
adopted here was to first determine these two parameters by using the
chiral angle of the classical Skyrme model, which is independent of both
model parameters and the representation~\cite{Norvaisas94}, so that
the empirical values of
the mass ($M_n =939$~MeV) and  isoscalar radius ($0.77$~fm)
of the nucleon are reproduced.
These parameters were then used in a numerical solution of
integrodifferential equation for the chiral angle~\cite{Acus98}, which
does
depend on the dimension of the representation
in the case of the quantized skyrmion. That solution was subsequently
used to determine new values of $f_\pi$ and $e$. This procedure was 
iterated until a convergent solution was obtained.

The values for the parameters found by this method are listed in
Table~\ref{t1} for a set of representations of the SU(2) group
with different dimension $j$. The chiral angle for the deuteron state, 
$F_R(r)$ was then
determined by self consistent numerical variation of the energy 
expression (\ref{mass}), for the representations listed in Table~\ref{t1}. 

The numerical results for the 
deuteron mass $M_d$, the binding energy $\Delta
E=M_d-2M_n$, matter radius $r_m$, charge radius $r_{ch}$, electric
quadrupole moment $Q_d$ and the mass $\tilde m$ (\ref{m2}),
which describes the falloff rate of the deuteron mass distribution at 
large distances, are listed in
Table~\ref{t1} for the irreducible representations with dimension $j=1/2,1$ and
$3/2$ as well as for the reducible representation
$1\oplus 1/2\oplus 1/2$. The quantum correction to the ground state
energy is similar in size to that found in ref.\cite{Leese} by
an entirely different approach.
The value of the falloff mass $\tilde m$,
is in all cases considered of the same order of magnitude as the 
quantum mechanical value $2\sqrt{B M_n}$, and in the case of the
three dimensional representation actually agrees with that value. The
value is also close in the case of the reducible representation.
It is interesting that these two representations are also those,
which give the best values for the falloff rate for the
matter density of the nucleon, which
should be of the order of the pion mass \cite{Acus98}.  

\begin{figure}
\begin{center}
\epsfig{file = 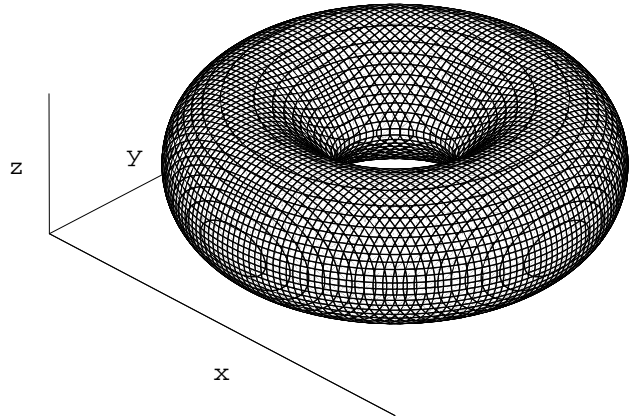}
\caption{Equidensity surface for the quantized deuteron}
\label{f1}
\end{center}
\end{figure}

\begin{figure}
\begin{center}
\epsfig{file = 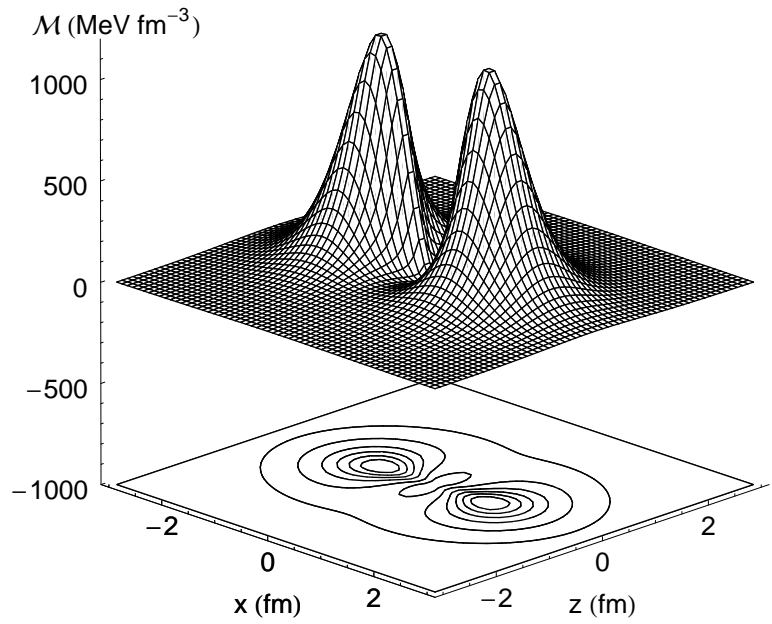}
\caption{Mass density distribution of the quantized deuteron}
\label{f2}
\end{center}
\end{figure}

The shape of deuteron is represented by the mass density
distribution $\mathcal{M} _{cl}+\Delta \mathcal{M}$. 
The equidensity surface displayed in Fig.~\ref{f1} is roughly toroidal.
The maximum value of mass density is 1150 MeV$\cdot$fm$^{-3}$ as shown
in Fig.~\ref{f2}. The density maxima form a ring with a
diameter of $1.424$~fm. The baryon density distribution has a similar shape.
An analogous toroidal structure in the deuteron has been shown to
arise in the quantum mechanical treatment of the two-nucleon
system with a realistic
interaction Hamiltonian \cite{Forest96}. 

\begin{figure}
\begin{center}
\epsfig{file = 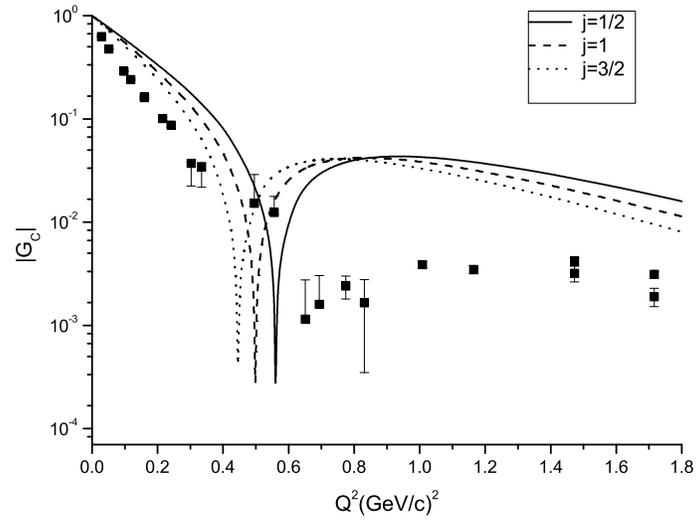}
\caption{Electric form factor of the quantized 
deuteron solution. The experimental data are
from \cite{Abbott2000}}
\label{f3}
\end{center}
\end{figure}

The calculated nonrelativistic charge form factor is shown in 
Fig.~\ref{f3} for the representations with dimension
$j=\frac12$, $1$ and $\frac32$. The calculated form factor
has the same qualitative features as the empirical form
factor value, although the charge radii are much too small.
The calculated charge and matter radii as well as the
quadrupole moments are listed in Table~\ref{t1}. These values
are only about half as large as the corresponding 
empirical values, a result which is similar to that
found for the semiclassical axisymmetric skyrmion
description of the deuteron. More realistic values for these
static parameters, which represent the large scale
features of the deuteron obtain with Skyrme's product
ansatz for the two-nucleon system \cite{Nyman87}.

\begin{acknowledgments}
Research supported in part by the Academy of Finland 
through grant 54038.
\end{acknowledgments}

\begin{table}
\caption{The predicted static deuteron observables in different
representations with fixed empirical values for nucleon isoscalar
radius $0.77$ fm. and mass $939$ MeV.}
\label{t1}
\begin{center}
\begin{tabular}{c|c|c|c|c|c}
\hline
$j$&$1/2$&$1$&$3/2$ &$\scriptscriptstyle 1\oplus \frac{1}{2} \oplus \frac12$&\rm{Expt.} \\ \hline
$f_\pi $& 57.68 & 56.53 & 55.71 & 56.82 &$93$ MeV \\ \hline
$e$ & 4.325 & 4.08 & 3.79 & 4.13 &  \\ \hline
$M_d$ & 1868 & 1926 & 1998 & 1907 & $1876$ MeV\\  \hline 
$\Delta E$ & $-$10 & 48 & 120 & 29 & $-2.22$ MeV\\ \hline
$r_{ch}$ &1.10&1.17&1.24&1.15&$2.13$ fm\\ \hline
$r_m$ &1.22&1.29&1.35&1.27&$1.97$ fm\\ \hline
$Q_d$ &0.140&0.158&0.177&0.152&$0.286$ fm${}^2$ \\  \hline
$m$ &54.6&90.5&110.0&82.3&$91.4$ MeV\\  \hline
\end{tabular}
\end{center}
\end{table}

\bibliographystyle{plain}

\newpage 

\end{document}